\documentclass{emulateapj}
\usepackage{amssymb}
\usepackage{amsmath}
\usepackage{graphicx}
\usepackage{subfigure}
\usepackage{wrapfig}
\usepackage[colorlinks,linkcolor=blue,anchorcolor=green,citecolor=blue]{hyperref}

\begin{document}

\def\func#1{\mathop{\rm #1}\nolimits}
\def\unit#1{\mathord{\thinspace\rm #1}}

\title{Optical transients powered by magnetars: dynamics, light curves, and
transition to the nebular phase}
\author{Ling-Jun Wang\altaffilmark{1,2,3}, S. Q. Wang\altaffilmark{2,3}, Z.
G. Dai\altaffilmark{2,3}, Dong Xu\altaffilmark{1}, Yan-Hui Han%
\altaffilmark{1}, X. F. Wu\altaffilmark{4,5}, Jian-Yan Wei\altaffilmark{1}}

\begin{abstract}
Millisecond magnetars can be formed via several channels: core-collapse of
massive stars, accretion-induced collapse of white dwarfs (WDs), double WD
mergers, double neutron star (NS) mergers, and WD-NS mergers. Because the
mass of ejecta from these channels could be quite different, their light
curves are also expected to be diverse. We evaluate the dynamic evolution of
optical transients powered by millisecond magnetars. We find that the
magnetar with short spin-down timescale converts its rotational energy
mostly into the kinetic energy of the transient, while the energy of a
magnetar with long spin-down timescale goes into radiation of the transient.
This leads us to speculate that hypernovae could be powered by magnetars
with short spin-down timescales. At late times the optical transients will
gradually evolve into a nebular phase because of the photospheric recession.
We treat the photosphere and nebula separately because their radiation
mechanisms are different. In some cases the ejecta could be light enough
that the magnetar can accelerate it to a relativistic speed. It is well
known that the peak luminosity of a supernova (SN) occurs when the
luminosity is equal to the instantaneous energy input rate, as shown by
Arnett (1979). We show that photospheric recession and relativistic motion
can modify this law. The photospheric recession always leads to a delay of
the peak time $t_{\mathrm{pk}}$ relative to the time $t_{\times }$ at which
the SN luminosity equals the instantaneous energy input rate. Relativistic
motion, however, may change this result significantly.
\end{abstract}



\affil{\altaffilmark{1}Key Laboratory of Space Astronomy and Technology, National
Astronomical Observatories, Chinese Academy of Sciences, Beijing 100012,
China; wanglj@nao.cas.cn, wjy@nao.cas.cn}

\affil{\altaffilmark{2}School of Astronomy and Space Science, Nanjing University, Nanjing,
China; dzg@nju.edu.cn}

\affil{\altaffilmark{3}Key laboratory of Modern Astronomy and Astrophysics (Nanjing
University), Ministry of Education, Nanjing 210093, China}

\affil{\altaffilmark{4}Purple Mountain Observatory, Chinese Academy of
Sciences, Nanjing, 210008, China}

\affil{\altaffilmark{5}Joint Center for Particle Nuclear Physics and Cosmology of Purple Mountain
Observatory-Nanjing University, Chinese Academy of Sciences, Nanjing 210008, China}

\section{Introduction}

The discovery of superluminous supernovae %
\citep[SLSNe;][]{Chomiuk11,Quimby2011,Gal-Yam12} in the last decade has
greatly expanded the SN family and its astrophysical significance, thanks to
the unbiased wide surveys such as the Panoramic Survey Telescope and Rapid
Response System (Pan-STARRS), the Palomar Transient Factory %
\citep[PTF;][]{Law09,Rau09}, the Catalina Realtime Transient Survey %
\citep[CRTS;][]{Drake09}, and the La Silla QUEST survey %
\citep[LSQ;][]{Baltay13}.

To date the observed SLSNe can be classified into two categories, viz. type
Ic SLSNe and type IIn SLSNe\footnote{%
A few SLSNe are classified as type IIL.}. Type IIn SLSNe show strong signs
of circumstellar interaction. Type Ic SLSNe, on the other hand, are usually
assumed to be energized by newborn millisecond magnetars during the core
collapse of the progenitors 
\citep{Kasen10,Woosley10,
Chatzopoulos12,Chatzopoulos13,Inserra13,McCrum14,Nicholl14,Papadopoulos15,
WangLiu15,WangWang15,Dai16,Metzger15,Mosta15}\footnote{%
It is suggested that the luminous supernovae show evidence for both $^{56}$%
Ni and magnetars \citep{WangWang2015b}.}.

Other channels to form millisecond magnetars include accretion-induced
collapse of white dwarfs \citep[AIC;][]{Canal76,Ergma76,Nomoto91,Usov92},
double white dwarf (WD) mergers \citep{Saio85,Levan06b}, double neutron star
mergers \citep[NSM;][]{dai98a,dai98b,dai06,zhang13,Giacomazzo13}, and WD-NS
mergers \citep{Metzger12}. AIC and NSM are of particular interest here
because AIC ejects an outflow with mass $M_{\mathrm{ej}}=10^{-3}-10^{-1}M_{%
\odot }$ \citep{Dessart06}, while NSM results in ejecta with masses $M_{%
\mathrm{ej}}=10^{-4}-10^{-2}M_{\odot }$ and velocity $v=0.1-0.3c$ %
\citep{rezzolla10,hotokezaka13,rosswog13}. Such low masses of the ejecta can
be accelerated to quasi-relativistic speed by the magnetar.

The magnetar-powered optical transients, e.g. a fraction of SLSNe that are
believed to be powered by magnetars, are usually modeled based on the
analytical solution developed by \cite{Arnett80,Arnett82}, who assumed that
the kinetic energy of the SNe is constant and the ejecta of the SNe are
dense enough that the photosphere does not recede. This simple solution
works pretty well in reproducing the observational data except for some
cases which render excess in theoretical light curves compared with the
observational data in late times \citep{Inserra13,Nicholl14}. To rescue the
magnetar model, it is arguably suggested \citep{WangWang15} that the energy
injected from the rotating magnetar could leak away from the ejecta based on
the observation that the spin-down luminosity of the magnetar could be
dominated by high energy gamma rays \citep{Caraveo14}.

The analytical model usually treats with the ejecta as a whole, even if the
SN goes into a nebular phase in late times. \cite{Arnett89} considered the
photospheric recession resulting from the ion recombination. In their model
the ejecta interior of the photosphere are continuously heated by the
radioactive decay of $^{56}$Ni and $^{56}$Co while the ejecta exterior are
cool and transparent that emanates no radiation. In this paper we consider
the photospheric recession and treat the ejecta within and outside of the
photosphere separately so that the SN can transit from the photospheric
phase into a nebular phase smoothly. Our aim at present is not to elaborate
a full-fledged model of the nebular phase of the SN. But rather, as a first
step, we would like to demonstrate how the nebular emission takes over the
photospheric emission. In future a more elaborated model can be developed
based on this model.

A significant fraction of the magnetar's rotational energy can be converted
into the SN kinetic energy. As a result, the acceleration of the SN by the
magnetar should be taken into consideration. In Section \ref{sec:SN} we
present our model, taking into account the dynamic evolution, photospheric
recession, nebular phase takeover. In the extreme case for very light
ejecta, the SN can be boosted to a quasi-relativistic speed. We present the
light curves and temperature evolution in Section \ref{sec:rel} in the
relativistic case. Finally, discussions and conclusions are given in Section %
\ref{sec:discussion}.

\section{An analytical SN model that treats photosphere and nebula separately%
}

\label{sec:SN}

We assume a constant density and homologous expansion of the SN. An ordinary
SN is heated by the $^{56}$Ni and $^{56}$Co decay %
\citep{Colgate69,Colgate80,Arnett80,Arnett82} 
\begin{equation}
L_{\mathrm{Ni}}\left( t\right) =M_{\mathrm{Ni}}\left[ \left( \epsilon _{%
\mathrm{Ni}}-\epsilon _{\mathrm{Co}}\right) e^{-t/\tau _{\mathrm{Ni}%
}}+\epsilon _{\mathrm{Co}}e^{-t/\tau _{\mathrm{Co}}}\right] ,
\end{equation}%
where $M_{\mathrm{Ni}}$ is the mass of $^{56}$Ni, $\epsilon _{\mathrm{Ni}%
}=3.9\times 10^{10}\unit{erg}\unit{g}^{-1}\unit{s}^{-1}$, $\epsilon _{%
\mathrm{Co}}=6.78\times 10^{9}\unit{erg}\unit{g}^{-1}\unit{s}^{-1}$, $\tau _{%
\mathrm{Ni}}$ and $\tau _{\mathrm{Co}}$ are the lifetime of $^{56}$Ni and $%
^{56}$Co, respectively. For a magnetar-powered SLSN, the energy source is
the spin-down luminosity of the magnetar \citep{Ostriker71}%
\begin{equation}
L_{\mathrm{mag}}\left( t\right) =\frac{E_{\mathrm{sd}}}{\tau _{\mathrm{sd}%
}\left( 1+t/\tau _{\mathrm{sd}}\right) ^{2}},
\end{equation}%
where $\tau _{\mathrm{sd}}=2.3\unit{days}R_{\ast
,6}^{-6}B_{p,14}^{-2}P_{0,-3}^{2}$ is the spin-down timescale of the
magnetar, $E_{\mathrm{sd}}=L_{\mathrm{sd},0}\tau _{\mathrm{sd}}$, $L_{%
\mathrm{sd},0}=10^{47}\unit{erg}\unit{s}^{-1}P_{0,-3}^{-4}B_{p,14}^{2}R_{%
\ast ,6}^{6}$ the spin-down luminosity of the magnetar, $B_{p}$ the
magnetar's dipole magnetic field, $P_{0}$ its initial spin period, $R_{\ast
} $ its radius. Here the usual convention $Q=10^{n}Q_{n}$ is adopted.

After taking into account the gamma-ray leakage, the radioactive decay input
within the photosphere is modified as%
\begin{equation}
L_{\mathrm{Ni}}^{\mathrm{SN}}\left( t\right) =L_{\mathrm{Ni}}\left( t\right)
\left( 1-e^{-\tau _{\gamma ,\mathrm{Ni}}^{\mathrm{SN}}}\right) ,
\end{equation}%
where the optical depth of the SN within photosphere to the $^{56}$Ni and $%
^{56}$Co decay photons $\tau _{\gamma ,\mathrm{Ni}}^{\mathrm{SN}}$ is given
by%
\begin{equation}
\tau _{\gamma ,\mathrm{Ni}}^{\mathrm{SN}}=\frac{3\kappa _{\gamma ,\mathrm{Ni}%
}M_{\mathrm{ej}}x_{\mathrm{ph}}}{4\pi R^{2}}.
\end{equation}%
Here $\kappa _{\gamma ,\mathrm{Ni}}$ is the opacity to the $^{56}$Ni and $%
^{56}$Co decay photons, $R\left( t\right) $ the SN radius (including the
nebula) at time $t$, and $x_{\mathrm{ph}}=v_{\mathrm{ph}}/v_{\mathrm{sc}}$. $%
v_{\mathrm{sc}}$ is the velocity scale which is approximately the
photospheric velocity $v_{\mathrm{ph}}$ at the SN peak luminosity when the
photosphere does not recede yet. In the homologous approximation, the
(relative) Lagrangian space coordinate of a fluid element is $x=r/R\left(
t\right) $.

Similarly, the spin-down luminosity taking into account the gamma-ray
leakage is%
\begin{equation}
L_{\mathrm{mag}}^{\mathrm{SN}}\left( t\right) =L_{\mathrm{mag}}\left(
t\right) \left( 1-e^{-\tau _{\gamma ,\mathrm{mag}}^{\mathrm{SN}}}\right) ,
\end{equation}%
where%
\begin{equation}
\tau _{\gamma ,\mathrm{mag}}^{\mathrm{SN}}=\frac{3\kappa _{\gamma ,\mathrm{%
mag}}M_{\mathrm{ej}}x_{\mathrm{ph}}}{4\pi R^{2}}
\end{equation}%
and $\kappa _{\gamma ,\mathrm{mag}}$ is the opacity to the gamma-ray photons
from the magnetar. Because the $^{56}$Ni gamma-ray spectrum is different
from that from the spinning-down magnetar, $\kappa _{\gamma ,\mathrm{mag}}$
is generally not the same as $\kappa _{\gamma ,\mathrm{Ni}}$. As can be
appreciated, the superscript {\small SN} in this paper indicates the
quantity within the SN photosphere, while the superscript {\small atm }in
the below context indicates the quantity within the SN nebula.

In the approximation of a constant expansion rate and that the initial
radius $R\left( 0\right) $ of the SN can be ignored, the radius of the SN is
given by $R=v_{\mathrm{sc}}t$. As a result the optical depth of the SN to
the gamma-rays within photosphere is given by%
\begin{equation}
\tau _{\gamma }^{\mathrm{SN}}=At^{-2},
\end{equation}%
where 
\begin{equation}
A=\frac{3\kappa _{\gamma }M_{\mathrm{ej}}}{4\pi v_{\mathrm{sc}}^{2}}.
\end{equation}%
Here $\kappa _{\gamma }$ can be $\kappa _{\gamma ,\mathrm{mag}}$ or $\kappa
_{\gamma ,\mathrm{Ni}}$, whatever suitable.

In the approximation that the photosphere does not recede, i.e. $x_{\mathrm{%
ph}}=1$, the luminosity is given by \citep{Arnett82}%
\begin{equation}
L\left( 1,t\right) =\frac{E_{\mathrm{th}}\left( 0\right) }{\tau _{0}}\phi
\left( t\right) ,  \label{eq:SN-L-no-recede}
\end{equation}%
where the \textquotedblleft 1" in $L\left( 1,t\right) $\ means the position $%
x=1$, i.e. the photosphere, and $E_{\mathrm{th}}\left( 0\right) $ is the
initial thermal energy of the SN and the diffusion time scale $\tau _{0}$ is
defined as \citep{Arnett80,Arnett82}%
\begin{equation}
\tau _{0}=\frac{\kappa M_{\mathrm{ej}}}{\tilde{\beta}cR\left( 0\right) },
\end{equation}%
where $\kappa $ is the opacity in optical band, $c$ the speed of light, and $%
\tilde{\beta}\simeq 13.8$. The function $\phi $, with initial value $\phi
\left( 0\right) =1$, evolves according to%
\begin{equation}
\dot{\phi}=\frac{R\left( t\right) }{R\left( 0\right) }\left[ \frac{L_{%
\mathrm{inp}}^{\mathrm{SN}}\left( t\right) }{E_{\mathrm{th}}\left( 0\right) }%
-\frac{\phi }{\tau _{0}}\right]  \label{eq:SN-phi-dot-no-recede}
\end{equation}%
where the energy input is the sum from $^{56}$Ni and magnetar%
\begin{equation}
L_{\mathrm{inp}}^{\mathrm{SN}}\left( t\right) =L_{\mathrm{Ni}}^{\mathrm{SN}%
}\left( t\right) +L_{\mathrm{mag}}^{\mathrm{SN}}\left( t\right) .
\end{equation}%
If the initial thermal energy of the SN can be neglected, we arrive at the
following integration expression of the luminosity\footnote{%
This is the equation used by \cite{Dai16}, which is different from the
widely used equation in the literature \citep[e.g.,][]{Chatzopoulos12},
which puts the gamma-ray leakage factor outside of the integration.}%
\begin{equation}
L(t)=\frac{2}{\tau _{m}}e^{-\frac{t^{2}}{\tau _{m}^{2}}}\int_{0}^{t}e^{\frac{%
t^{\prime 2}}{\tau _{m}^{2}}}\frac{t^{\prime }}{\tau _{m}}L_{\mathrm{inp}%
}(t^{\prime })\left[ 1-e^{-\tau _{\gamma }(t^{\prime })}\right] dt^{\prime },
\end{equation}%
where the effective diffusion timescale $\tau _{m}$ is given by%
\begin{equation}
\tau _{m}=\left( \frac{2\kappa M_{\mathrm{ej}}}{\tilde{\beta}cv_{\mathrm{sc}}%
}\right) ^{1/2}.
\end{equation}

Generally the photosphere will steadily recede, in this case the
photospheric emission is given by \citep{Arnett89}%
\begin{equation}
L_{\mathrm{ph}}\left( x_{\mathrm{ph}},t\right) =\frac{E_{\mathrm{th}}\left(
0\right) }{\tau _{0}}x_{\mathrm{ph}}\phi ,  \label{eq:SN-ph-lum}
\end{equation}%
where $\phi \left( t\right) $ evolves as \citep{Arnett89}\footnote{%
Equation $\left( \text{A41}\right) $ in \cite{Arnett89} is in error. The
correct one should read 
\begin{equation*}
\frac{d\phi }{dz}=\frac{\sigma }{x_{i}^{3}}\left[ p_{1}\zeta \left( t\right)
-p_{2}x_{i}\phi -3x_{i}^{2}\frac{\phi }{\sigma }\frac{dx_{i}}{dz}\right] ,
\end{equation*}%
where $p_{1}$, $p_{2}$ are defined by Equations (A20) and (A21) in \cite%
{Arnett89}, respectively.}%
\begin{equation}
\dot{\phi}=\frac{1}{x_{\mathrm{ph}}^{3}}\frac{R\left( t\right) }{R\left(
0\right) }\left[ \frac{L_{\mathrm{inp}}^{\mathrm{SN}}\left( t\right) }{E_{%
\mathrm{th}}\left( 0\right) }-\frac{x_{\mathrm{ph}}}{\tau _{0}}\phi -3x_{%
\mathrm{ph}}^{2}\frac{R\left( 0\right) }{R\left( t\right) }\dot{x}_{\mathrm{%
ph}}\phi \right] ,  \label{eq:phi_dot_photosphere}
\end{equation}%
where we have substituted $x_{i}$ in Equation $\left( \text{A41}\right) $ of 
\cite{Arnett89} for $x_{\mathrm{ph}}$. The Lagrangian coordinate of the
photosphere $x_{\mathrm{ph}}$ is given by%
\begin{equation}
x_{\mathrm{ph}}=1-\frac{2}{3}\frac{\lambda }{R\left( t\right) },
\end{equation}%
where $\lambda =\left( \rho \kappa \right) ^{-1}$ is the mean free path of
the SN photons and $\rho $ the SN material density. The effective
temperature of the photosphere is%
\begin{equation}
T_{\mathrm{ph}}^{4}\left( t\right) =\frac{E_{\mathrm{th}}\left( 0\right) }{%
4\pi R^{2}\sigma \tau _{0}}\frac{\phi }{x_{\mathrm{ph}}}.
\end{equation}

Up to now we have not specified the evolution of the scale velocity $v_{%
\mathrm{sc}}$. Under the assumption of homologous expansion of the SN, the
kinetic energy of the SN is given by \citep{Arnett82}%
\begin{equation}
E_{\mathrm{SN}}=\frac{3}{10}M_{\mathrm{ej}}v_{\mathrm{sc}}^{2},
\end{equation}%
whereby the scale velocity can be determined as%
\begin{equation}
v_{\mathrm{sc}}=\left[ \left( \frac{5}{3}\right) \frac{2\left( E_{\mathrm{SN}%
,0}+E_{K,\mathrm{inp}}\right) }{M_{\mathrm{ej}}}\right] ^{1/2}
\end{equation}%
Here $E_{\mathrm{SN},0}$ is the initial kinetic energy of the SN and the
kinetic energy input $E_{K,\mathrm{inp}}$ is determined by%
\begin{equation}
\frac{dE_{K,\mathrm{inp}}}{dt}=L_{K}-L_{\mathrm{ph}}-L_{\mathrm{atm}}.
\label{eq:E_K_evolve}
\end{equation}%
The energy input from the magnetar, $L_{K}$, is given by%
\begin{equation}
L_{K}=L_{\mathrm{mag}}\left( t\right) \left( 1-e^{-\tau _{\gamma ,\mathrm{mag%
}}}\right) ,
\end{equation}%
where the optical depth of the SN (including the nebula) to the gamma-ray
from the magnetar is given by%
\begin{equation}
\tau _{\gamma ,\mathrm{mag}}=\frac{3\kappa _{\gamma ,\mathrm{mag}}M_{\mathrm{%
ej}}}{4\pi R^{2}}.
\end{equation}%
The nebular luminosity, $L_{\mathrm{atm}}$, in Equation $\left( \ref%
{eq:E_K_evolve}\right) $ will be determined as follows.

For the nebular component of the SN, i.e. the part outside of the
photosphere of the SN, we assume a homogeneous density and temperature
distribution. In this approximation, the nebular luminosity is%
\begin{equation}
L_{\mathrm{atm}}=\frac{E_{\mathrm{atm}}c}{R\left( 1-x_{\mathrm{ph}}\right) },
\end{equation}%
where the internal energy of the nebula evolves according to%
\begin{equation}
\frac{dE_{\mathrm{atm}}}{dt}+P\frac{dV_{\mathrm{atm}}}{dt}=L_{\mathrm{inp}}^{%
\mathrm{atm}}-L_{\mathrm{atm}}-4\pi R^{3}x_{\mathrm{ph}}^{2}aT_{\mathrm{ph}%
}^{4}\frac{dx_{\mathrm{ph}}}{dt}.  \label{eq:E_atm_evolve}
\end{equation}%
Obviously, the volume of the nebula is%
\begin{equation}
V_{\mathrm{atm}}=\frac{4}{3}\pi R^{3}\left( 1-x_{\mathrm{ph}}^{3}\right) .
\end{equation}%
In Equation $\left( \ref{eq:E_atm_evolve}\right) $ the first term on the
right hand side is the energy input from magnetar and $^{56}$Ni%
\begin{equation}
L_{\mathrm{inp}}^{\mathrm{atm}}=L_{\mathrm{Ni}}^{\mathrm{atm}}\left(
t\right) +L_{\mathrm{mag}}^{\mathrm{atm}}\left( t\right) ,
\end{equation}%
which is the sum of the following two terms%
\begin{eqnarray}
L_{\mathrm{Ni}}^{\mathrm{atm}}\left( t\right) &=&L_{\mathrm{Ni}}\left(
t\right) \left( e^{-\tau _{\gamma ,\mathrm{Ni}}^{\mathrm{SN}}}-e^{-\tau
_{\gamma ,\mathrm{Ni}}}\right) , \\
L_{\mathrm{mag}}^{\mathrm{atm}}\left( t\right) &=&L_{\mathrm{mag}}\left(
t\right) \left( e^{-\tau _{\gamma ,\mathrm{mag}}^{\mathrm{SN}}}-e^{-\tau
_{\gamma ,\mathrm{mag}}}\right) .
\end{eqnarray}%
Finally the nebular temperature is%
\begin{equation}
T_{\mathrm{atm}}=\left( \frac{E_{\mathrm{atm}}}{aV_{\mathrm{atm}}}\right)
^{1/4}.
\end{equation}

Having elaborated our analytical model, we would like to evaluate the model
against observational data. First of all, because the magnetar can deposit a
significant fraction of its rotational energy as the kinetic energy of the
SN, the expansion velocity of the SN is actually not a constant, as is
usually assumed in the previous analytical models. However, since the
magnetar dissipates its rotational energy in a timescale $\tau _{\mathrm{sd}%
}=2.3\unit{days}R_{6}^{-6}B_{p,14}^{-2}P_{0,-3}^{2}$ the boost of the SN
kinetic energy by magnetar cannot be observed given the fact that the
typical SNe are detected several days after their explosion when the
magnetars have already exhausted their rotational energy.

\begin{figure*}[tbph]
\centering\includegraphics[width=0.33\textwidth,angle=-90]{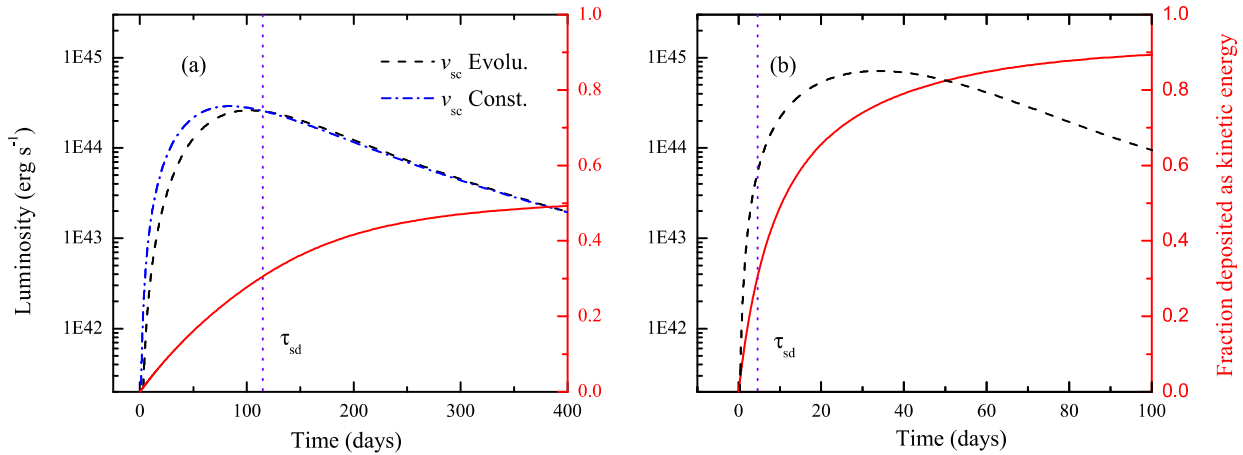}
\caption{SN light curves (dashed lines) for the parameters $M_{\mathrm{ej}%
}=10M_{\odot }$, $B_{p}=2\times 10^{13}\unit{G}$, $P_{0}=1.4\unit{ms}$
(panel a, i.e. the case $\protect\tau _{\mathrm{sd}}\gtrsim \protect\tau %
_{m} $) and $M_{\mathrm{ej}}=10M_{\odot }$, $B_{p}=10^{14}\unit{G}$, $%
P_{0}=1.4\unit{ms}$ (panel b, i.e. the case $\protect\tau _{\mathrm{sd}}\ll 
\protect\tau _{m}$). The solid lines (right vertical axis) show the
accumulative fraction of the rotational energy of the magnetar converted to
the kinetic energy of the SN. Vertical dotted lines mark the time $\protect%
\tau _{\mathrm{sd}}$. Panel (a) also compares the SN light curves calculated
by taking into account the acceleration of SN by the magnetar (dashed line)
and that expected in a constant-velocity model (dot-dashed line). Note that
for rendering clarity, the abscissa time internals are different between
these two panels.}
\label{fig:LC-v}
\end{figure*}

The SN light curve rises to its maximum roughly in the effective diffusion
timescale $\tau _{m}$. To clearly observe the acceleration of the SN by
magnetar, the condition $\tau _{\mathrm{sd}}\gtrsim \tau _{m}$ should be
guaranteed, which leads to the following upper bound on the magnetar dipole
field strength%
\begin{equation}
B_{p,14}\lesssim 0.25R_{6}^{-3}P_{0,-3}\left( \frac{v_{\mathrm{sc},9}}{%
\kappa _{-1}M_{\mathrm{ej},1}}\right) ^{1/4},  \label{eq:B_upper_bound-i-e}
\end{equation}%
where the typical parameters $v_{\mathrm{sc}}=10^{9}\unit{cm}\unit{s}^{-1}$, 
$\kappa =0.1\unit{cm}^{2}\unit{g}^{-1}$, $M_{\mathrm{ej}}=10M_{\odot }$ are
adopted. In the above expression we assume that the initial explosion energy
of the SN, $E_{\mathrm{SN}}$, is larger than the energy deposited as kinetic
energy by the magnetar. This is however not always true. If we assume that a
fraction $\eta _{K}$ of the rotational energy of the magnetar can be
deposited as the kinetic energy of the SN, the magnetar along can accelerate
the SN to a velocity%
\begin{equation}
v_{\mathrm{sc}}=1.8\times 10^{9}\unit{cm}\unit{s}^{-1}\eta
_{K}^{1/2}I_{45}^{1/2}M_{\mathrm{ej},1}^{-1/2}P_{0,-3}^{-1},
\end{equation}%
where $I$ is the moment of inertia of the magnetar. Obviously, for less
massive ejecta, the SN can be accelerated to high speed by a rapidly
rotating magnetar. Provided that the kinetic energy of the SN is dominantly
attributed to the magnetar, to fulfil the condition $\tau _{\mathrm{sd}%
}\gtrsim \tau _{m}$, we have%
\begin{equation}
B_{p,14}\lesssim 0.26R_{6}^{-3}P_{0,-3}^{3/4}\kappa _{-1}^{-1/4}M_{\mathrm{ej%
},1}^{-3/8}.  \label{eq:B_upper_bound-mag}
\end{equation}%
We find that Equations $\left( \ref{eq:B_upper_bound-i-e}\right) $ and $%
\left( \ref{eq:B_upper_bound-mag}\right) $ give similar upper bounds on the
dipole magnetic field.

In Figure \ref{fig:LC-v}(a) we show the effect of the kinetic energy
injection by the magnetar, along with the accumulative fraction of the total
rotational energy of the magnetar deposited as the kinetic energy of the SN.
In Figure \ref{fig:LC-v}(a) the following values are adopted $B_{p}=2\times
10^{13}\unit{G}$, $M_{\mathrm{ej}}=10M_{\odot }$, $P_{0}=1.4\unit{ms}$. In
appreciating the kinetic injection effect, we set the asymptotic velocity in
our model to equal to the constant velocity in the usual model. It is clear
from Figure \ref{fig:LC-v}(a) that the kinetic injection affects the rising
part of the light curve, with a luminosity dimmer than the constant velocity
model. Figure \ref{fig:LC-v}(a) also shows that at the maximum $\eta
_{K}\sim 30\%$ of the rotational energy of the magnetar is deposited as the
kinetic energy of the SN, in rough agreement with that determined by \cite%
{Woosley10}. But ultimately this fraction can be as high as $\eta _{K}\sim
50\%$. In Figure \ref{fig:LC-v}(a) the parameters are adopted so that the
condition $\tau _{\mathrm{sd}}>\tau _{m}$ is fulfilled. In the opposite
extreme, i.e. $\tau _{\mathrm{sd}}\ll \tau _{m}$, however, as can be seen
from Figure \ref{fig:LC-v}(b), as high as $\sim 100\%$ of the rotational
energy of the magnetar is converted into the kinetic energy of the SN. In
conclusion, we expect that in a magnetar-powered SN, $\sim 50\%-\left. \sim
100\%\right. $ of the rotational energy of the magnetar can be converted
into the kinetic energy of the SN.

In this model because $v_{\mathrm{sc}}$\ is no longer a constant, the
effective diffusion time $\tau _{m}$\ is not well defined. However, since
the variation of $v_{\mathrm{sc}}$\ is slow, we can still think of $\tau
_{m} $\ as the rising time of the light curve. Please note that the rising
times of the light curves in the two panels of Figure \ref{fig:LC-v} are
different even other parameters are the same except the dipole magnetic
field $B_{p}$. This is because the magnetar with a stronger magnetic field
can accelerate the ejecta to higher speed, reducing the effective diffusion
time $\tau _{m}$. In short, a magnetar with a short spin-down time deposits
its rotational energy mainly as the kinetic energy of the SN, while the one
with a long spin-down time deposits its rotational energy mainly as the
thermal radiation of the SN. We therefore conclude that for an SN to become
an SLSN, the magnetar must be fast rotating with a relatively weak dipole
magnetic field.

\begin{figure}[tbph]
\centering\includegraphics[width=0.48\textwidth,angle=0]{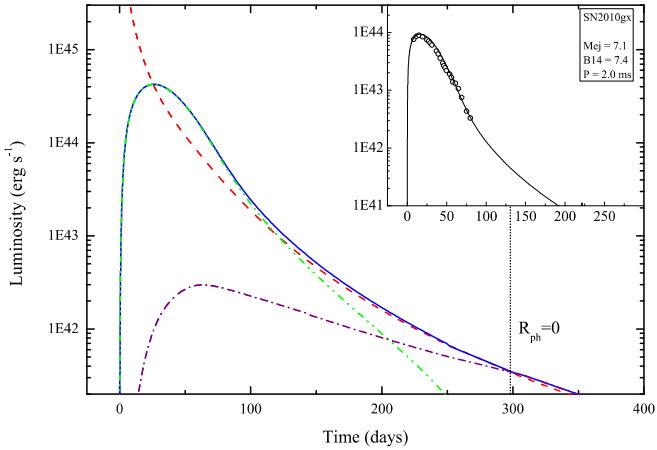}
\caption{The light curves calculated according to our model with parameters $%
M_{\mathrm{ej}}=10M_{\odot }$, $B_{p}=2\times 10^{14}\unit{G}$, $P_{0}=1.4%
\unit{ms}$. In this figure we shows the bolometric luminosity (solid line),
luminosity due to the photospheric emission (dot-dot-dashed line),
luminosity due to the nebula (dot-dashed line), and the magnetar luminosity
(dashed line). Here the energy leakage of the magnetar is taken into
account. The vertical dotted line marks the time when the photosphere of the
SN disappears. The inset is the fit of our model to the observational data
of SN2010gx, with the fit parameters the same as that used by \protect\cite%
{Inserra13}.}
\label{fig:SN-LC}
\end{figure}

In the following calculations of this Section, we take the magnetar
parameters $B_{p,14}=2$, $P_{0,-3}=1.4$, which is equivalent to $L_{\mathrm{%
sd},0}=10^{47}\unit{erg}\unit{s}^{-1}$ and $\tau _{\mathrm{sd}}=10^{5}\unit{s%
}$. Other parameters are $M_{\mathrm{ej}}=10M_{\odot }$, $\kappa =0.1\unit{cm%
}^{2}\unit{g}^{-1}$, $\kappa _{\gamma ,\mathrm{mag}}=0.02\unit{cm}^{2}\unit{g%
}^{-1}$, $M_{\mathrm{Ni}}=0$. We set the initial explosion energy of the SN
to be $E_{\mathrm{SN}}=0$.

\begin{figure}[tbph]
\centering\includegraphics[width=0.48\textwidth,angle=0]{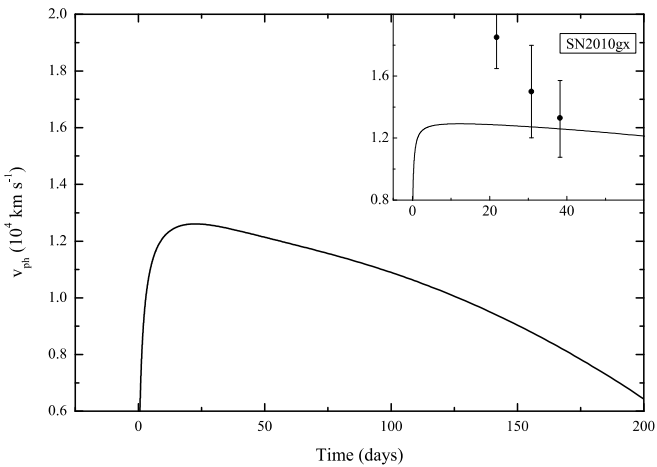}
\caption{Photospheric velocity for the model parameters the same as Figure 
\protect\ref{fig:SN-LC}. The inset is the velocity evolution of SN2010gx.}
\label{fig:SN-v}
\end{figure}

Figure \ref{fig:SN-LC} shows our model light curves. A particular
demonstration of the model in reproducing the observational data is
exemplified in the inset. We see that at early times the photospheric
emission completely dominates, whereas at late times the photospheric
emission and nebular emission conspire to follow the input power of the
magnetar. In this calculation the energy leakage of the magnetar is taken
into account. When $R_{\mathrm{ph}}=0$, $L_{\mathrm{atm}}=L_{\mathrm{inp}}^{%
\mathrm{mag}}$.

\begin{figure}[tbph]
\centering\includegraphics[width=0.48\textwidth,angle=0]{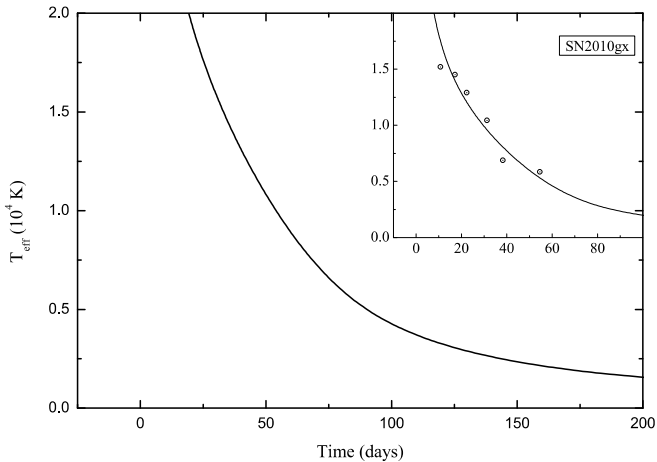}
\caption{Effective temperature for the model parameters the same as Figure 
\protect\ref{fig:SN-LC}. The inset is the temperature evolution of SN2010gx.}
\label{fig:SN-T}
\end{figure}

We show the photospheric velocity and effective temperature in Figures \ref%
{fig:SN-v} and \ref{fig:SN-T}, respectively. The effective temperature is a
weighted average of the photospheric emission and nebular emission, given by

\begin{equation}
T_{\mathrm{eff}}=\left( \frac{L_{\mathrm{ph}}T_{\mathrm{ph}}^{4}+L_{\mathrm{%
atm}}T_{\mathrm{atm}}^{4}}{L_{\mathrm{ph}}+L_{\mathrm{atm}}}\right) ^{1/4}.
\label{eq:weight_temp}
\end{equation}%
The above averaging scheme is motivated to pick out the temperature of the
dominant emission component and at the same time make a smooth transition
even if one of the components goes to zero.

\section{Relativistic motion}

\label{sec:rel}

In this section we assume that the ejecta are light enough, i.e. $M_{\mathrm{%
ej}}=10^{-4}-10^{-2}M_{\odot }$, so that the magnetar can accelerate it to
quasi-relativistic speed. Assuming the formation of a rapidly rotating
magnetar following the double neutron star merger, \cite{yu13} studied the
merger remnant emission, dubbed merger-nova, powered by a magnetar. However,
owing to some simplified assumptions underlining the calculations by \cite%
{yu13}, a revisit of this problem is necessary.

The merger-nova emission, in the comoving frame, can be obtained by the
method outlined in the above Section. To determine the photospheric radius
of the merger-nova, we need a knowledge of the opacity. \cite{yu13} adopted
the `canonical' value $\kappa =0.2\unit{cm}^{2}\unit{g}^{-1}$, whereas we
will take the newly determined value $\kappa =10\unit{cm}^{2}\unit{g}^{-1}$ %
\citep{barnes13,kasen13,tanaka13,grossman14}.

\cite{yu13} also takes into account the deceleration of the remnant by the
swept-up ambient media. We assume an ambient hydrogen number density $n=0.1%
\unit{cm}^{-3}$, as determined in the literature %
\citep{berger05,soderberg06,berger07,wang13b}. As a result, the ejecta
dynamics is determined by \citep{yu13}%
\begin{equation}
{\small \frac{d\Gamma }{dt}=\frac{\xi L_{\mathrm{sd}}+L_{\mathrm{ra}%
}-L_{e}-\Gamma \mathcal{D}\left( dE_{\mathrm{int}}^{\prime }/dt^{\prime
}\right) -\left( \Gamma ^{2}-1\right) c^{2}\left( dM_{\mathrm{sw}}/dt\right) 
}{M_{\mathrm{ej}}c^{2}+E_{\mathrm{int}}^{\prime }+2\Gamma M_{\mathrm{sw}%
}c^{2}},}  \label{eq:Gamma_dot}
\end{equation}%
based on the generic dynamic model of GRBs \citep{huang99}. Here the
quantities in the comoving frame are denoted by a prime. In the above
equation $E_{\mathrm{int}}^{\prime }$ is the internal energy of the ejecta
in the comoving frame, $\Gamma $ the Lorentz factor of the ejecta, $M_{%
\mathrm{sw}}$ the ambient mass swept up by the expanding merger-nova, $L_{e}$
the bolometric luminosity of the merger-nova, $t$ the time measured in the
observer frame. The fraction of the Poynting flux caught by the ejecta is
set as $\xi =0.8$. The magnetar emission leakage is taken into account by
setting $\kappa _{\gamma ,\mathrm{mag}}=0.02\unit{cm}^{2}\unit{g}^{-1}$, the
same as used in the above Section. The radius $R\left( t^{\prime }\right) $
in Equation $\left( \ref{eq:phi_dot_photosphere}\right) $ can be obtained by
a combination of Equation $\left( \ref{eq:Gamma_dot}\right) $ and%
\begin{equation}
\frac{dR}{dt}=\frac{\beta c}{1-\beta },
\end{equation}%
where $\beta =v/c$ is the dimensionless velocity of the ejecta front.

\begin{figure}[tbph]
\centering\includegraphics[width=0.4\textwidth,angle=-90]{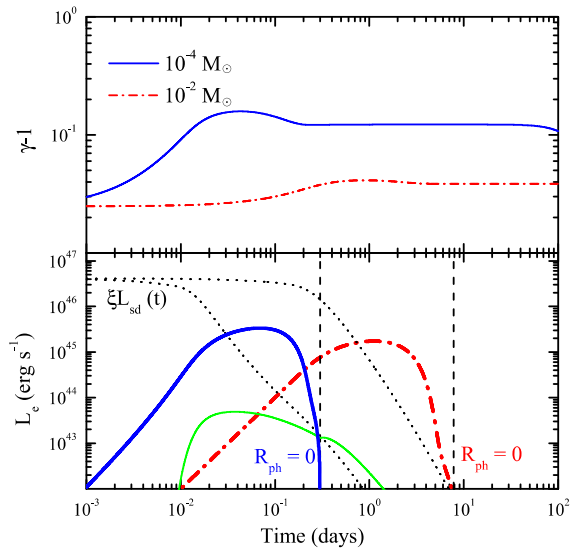}
\caption{The upper panel shows the evolution of Lorentz factor. The lower
panel shows the emission of photosphere (thick lines) and nebula (thin
lines). The nebular emission for the $M_{\mathrm{ej}}=10^{-2}M_{\odot }$
case is too low to be shown here. The dotted line is the spin down
luminosity caught up by the ejecta with energy leakage considered. The
vertical dashed lines mark the times when the photospheric radii vanish. The
model parameters are $B_{p,14}=2$, $P_{0,-3}=1.4$, $\protect\kappa =10\unit{%
cm}^{2}\unit{g}^{-1}$, $\protect\kappa _{\protect\gamma ,\mathrm{mag}}=0.02%
\unit{cm}^{2}\unit{g}^{-1}$.}
\label{fig:luminosity}
\end{figure}

In the comoving frame, $L_{\mathrm{mag}}^{\prime }=L_{\mathrm{mag}}/\mathcal{%
D}^{2}$, where $\mathcal{D}$ is the Doppler factor. In the following
calculations the mass and initial velocity of the ejecta are taken to be $M_{%
\mathrm{ej}}=10^{-4}$, $10^{-2}M_{\odot }$ and $v=0.2c$, in accordance with
the numerical simulations \citep{rezzolla10,hotokezaka13,rosswog13}.

In the comoving frame, the radioactive heating rate, $L_{\mathrm{ra}%
}^{\prime }$, of the $r$-process material can be well approximated according
to \citep{korobkin12}%
\begin{equation}
\dot{\epsilon}\left( t^{\prime }\right) =\epsilon _{0}\left( \frac{1}{2}-%
\frac{1}{{\pi }}\arctan \frac{t^{\prime }-t_{0}}{\sigma }\right)
^{1.3}\left( \frac{\epsilon _{\mathrm{th}}}{0.5}\right)
\label{eq:radio-heating-rate}
\end{equation}%
where $\epsilon _{\mathrm{th}}$ is the heating efficiency, and $\epsilon
_{0}=2\times 10^{18}\unit{erg}\unit{g}^{-1}\unit{s}^{-1}$, $t_{0}=1.3\unit{s}
$, and $\sigma =0.11\unit{s}$, $t^{\prime }$ the time measured in comoving
frame. The heating efficiency in Equation $\left( \ref{eq:radio-heating-rate}%
\right) $ is set as $\epsilon _{\mathrm{th}}=0.6$ to account for the energy
carried away by neutrinos and the energy leak caused by gamma-ray diffusion %
\citep{metzger10a,roberts11}. For the typical values of the ejecta mass, the
radioactive heating rate is negligible compared with the spin-down
luminosity of the magnetar.

\begin{figure}[tbph]
\centering\includegraphics[width=0.4\textwidth,angle=-90]{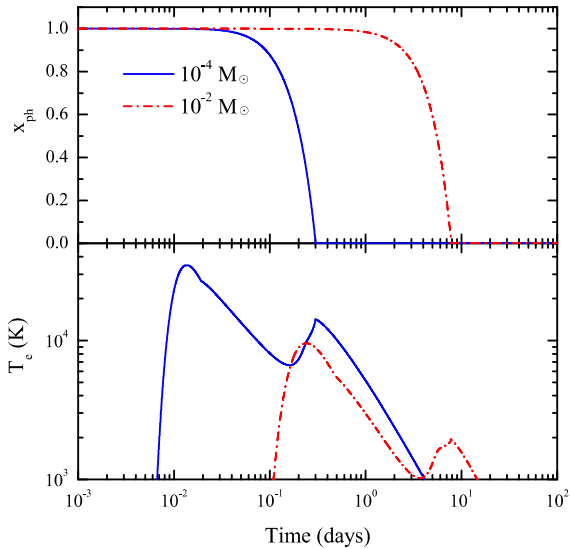}
\caption{The evolution of $x_{\mathrm{ph}}$ (upper panel) and effective
temperature in observer's frame (lower panel). The model parameters are the
same as Figure \protect\ref{fig:luminosity}.}
\label{fig:Te}
\end{figure}

The photospheric luminosity and flux density in the observer frame is given
by $L_{\mathrm{ph}}=L_{\mathrm{ph}}^{\prime }\mathcal{D}^{2}$ and%
\begin{equation}
F_{\nu }=\frac{2\pi \mathcal{D}^{2}R^{2}x_{\mathrm{ph}}^{2}}{%
D_{L}^{2}h^{3}c^{2}\nu }\frac{\left( h\nu /\mathcal{D}\right) ^{4}}{\exp
\left( h\nu /\mathcal{D}kT_{e}^{\prime }\right) -1},
\end{equation}%
respectively. Here $D_{L}$ is the distance of the merger-nova.

In Figure \ref{fig:luminosity} we show the Lorentz factor and bolometric
luminosities. We find that when considering the gamma-ray leakage, the
ejecta can only be accelerated to sub-relativistic speed. The Lorentz factor
experiences an increase and then a decline before the time when $R_{\mathrm{%
ph}}=0$. This is because the merger-nova luminosity varies from a value
below to a value above the magnetar input luminosity. The deceleration of
the ejecta at time $t\simeq 100\unit{days}$ is due to the impact of the
ambient medium. In the calculations in this Section, the model parameters
are $B_{p,14}=2$, $P_{0,-3}=1.4$, $\kappa =10\unit{cm}^{2}\unit{g}^{-1}$, $%
\kappa _{\gamma ,\mathrm{mag}}=0.02\unit{cm}^{2}\unit{g}^{-1}$.

Figure \ref{fig:Te} depicts evolution of the photospheric radius $x_{\mathrm{%
ph}}$ and the effective temperature $T_{e}$, weighted average expressed in $%
\left( \ref{eq:weight_temp}\right) $, in the observer frame. Because the
ejecta are barely relativistic, the effective temperature in the comoving
frame would be very close to the observed temperature. It is found that the
temperature does not monotonically decline after the first rapid rise. The
second rise occurs just before the photosphere disappears. This bebaviour of
the temperature evolution can be understood by comparing Figures \ref%
{fig:SN-LC} and \ref{fig:luminosity}. Figure \ref{fig:SN-LC} shows that long
before $R_{\mathrm{ph}}=0$, the nebular emission dominates over the
photospheric emission, which is in sharp contrast to the situation in Figure %
\ref{fig:luminosity}. In the quasi-relativistic case, the photospheric
emission dominates over the nebular emission before $R_{\mathrm{ph}}=0$, and
therefore the observed temperature reflects the photospheric emission. In
this case the photosphere recedes so rapidly that the inner hot part is
exposed to the observer.

\begin{figure}[tbph]
\centering\includegraphics[width=0.4\textwidth,angle=-90]{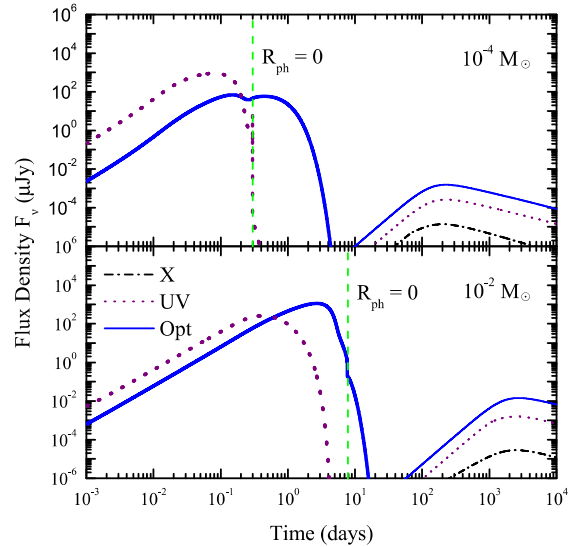}
\caption{Transient emission (thick lines) and forward shock emission (thin
lines) in X-ray ($1\unit{keV}$), UV ($30\unit{eV}$), and optical ($606\unit{%
nm}$) bands. The flux densities are calculated by assuming a luminosity
distance $D_{L}=10^{27}\unit{cm}$. Vertical dashed lines mark the times when
the photospheric radii vanish. The model parameters are the same as Figure 
\protect\ref{fig:luminosity}.}
\label{fig:photo-emission}
\end{figure}

In Figure \ref{fig:photo-emission} we show the observed flux densities of
the merger-nova emission (before $\sim 10\unit{days}$) and forward shock
emission (after $\sim 10\unit{days}$) at optical, UV, and X-ray bands. The
merger-nova emission before the time when $R_{\mathrm{ph}}=0$ comes
dominantly from photosphere, thereafter the emission comes from the nebula.
It is clear from this figure that the X-ray emission from merger-nova is
negligible.

\section{Discussions and Conclusions}

\label{sec:discussion}

In this paper, we develop a model for the magnetar-powered optical
transients to treat the photospheric and nebular emission separately based
on \cite{Arnett89}. We also consider the quasi-relativistic merger-nova. We
evaluate the acceleration of the SLSNe by the magnetar and find that this
effect can be observed most clearly in the case of $\tau _{\mathrm{sd}}>\tau
_{m}$, where $\sim 50\%$ of the magnetar rotational energy is converted to
the kinetic energy of the SLSNe. In the other extreme case, i.e. $\tau _{%
\mathrm{sd}}\ll \tau _{m}$, up to $100\%$ of the magnetar energy can be
deposited as the SN kinetic energy.

\begin{figure*}[tbph]
\centering\includegraphics[width=0.65\textwidth,angle=-90]{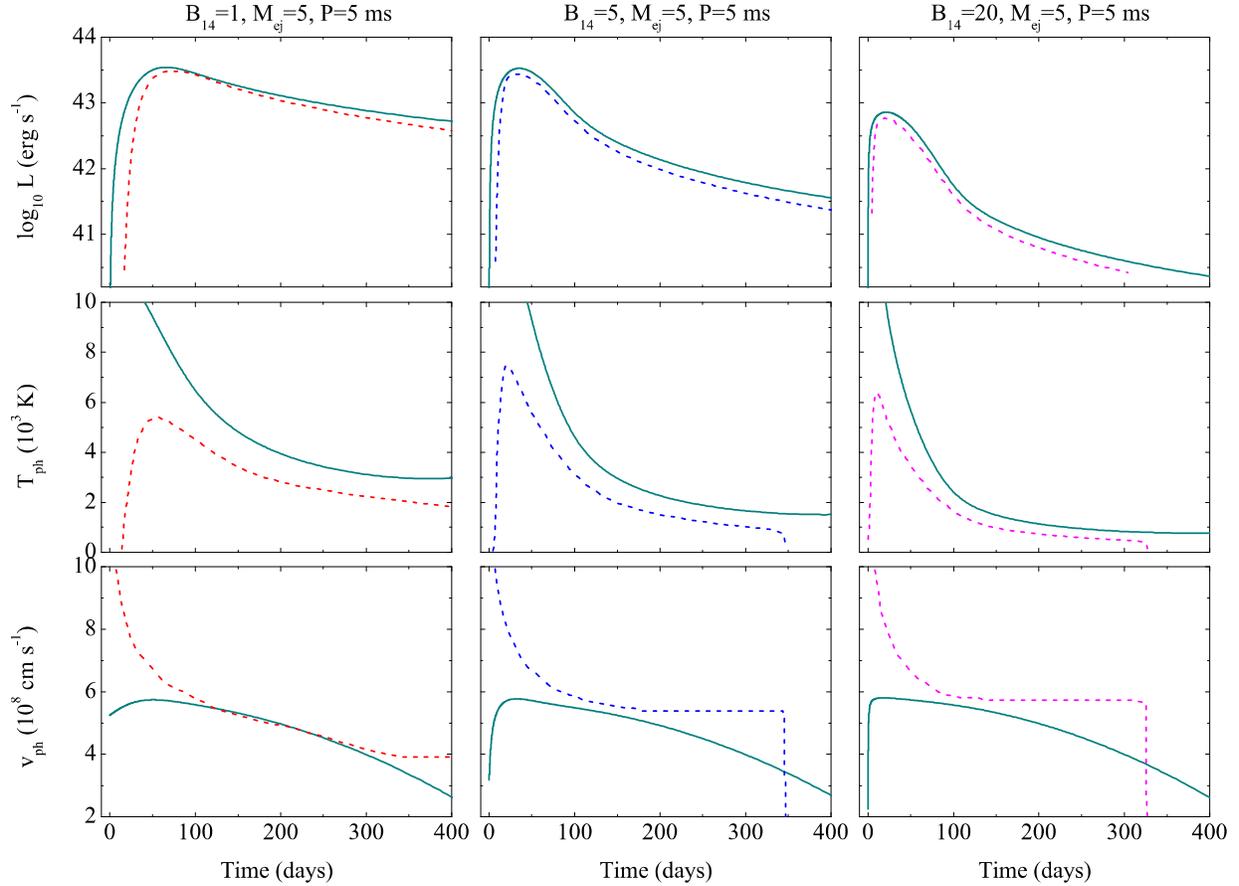}
\caption{Comparison of light curves, photospheric temperatures, and
photospheric velocities from our model (solid lines) with the hydrodynamical
simulations (dashed lines) of \protect\cite{Kasen10}. In this figure the
opacities are set as $\protect\kappa =0.2\unit{cm}^{2}\unit{g}^{-1}$, and $%
\protect\kappa _{\protect\gamma }=\infty $, in accord with \protect\cite%
{Kasen10}.}
\label{fig:Kasen}
\end{figure*}

It is found that the SLSNe transit from photospheric to nebular phase
smoothly. When $R_{\mathrm{ph}}=0$ the nebular luminosity is equal to the
instantaneous energy input rate of the magnetar. This is true for both
sub-relativistic SLSNe (Figure \ref{fig:SN-LC}) and quasi-relativistic
merger-novae (Figure \ref{fig:luminosity}). Nevertheless we note that in
this model we do not account for changes in ionization balance and hence in
opacity, which could in principle affect the nebular transition.

\begin{figure*}[tbph]
\centering\includegraphics[width=0.45\textwidth,angle=-90]{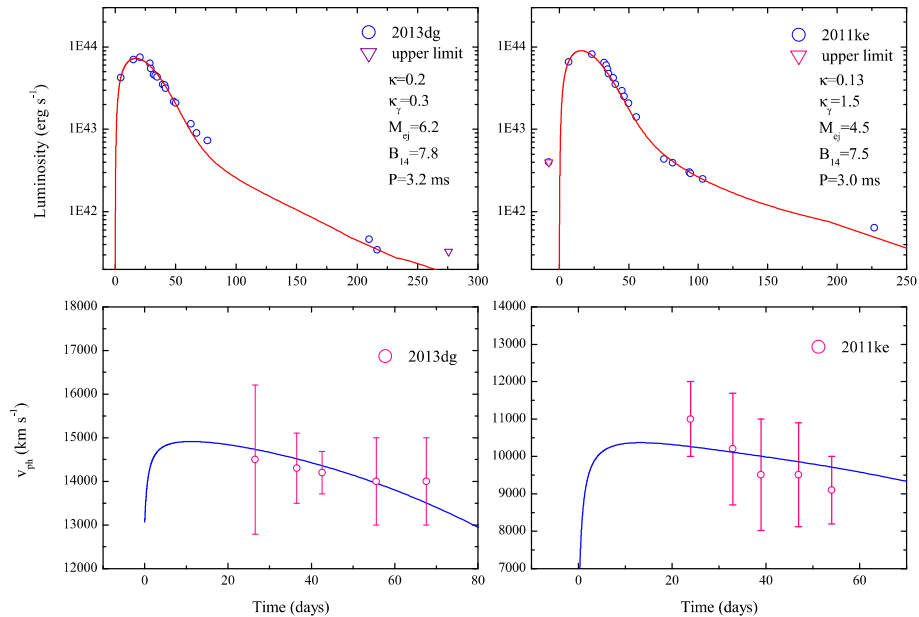}
\caption{A demonstration of our analytical model to account for the flat
velocity data points observed in most of the SLSNe.\ The light curve data
for SNe 2011ke and 2013dg are taken from \protect\cite{Inserra13} and 
\protect\cite{Nicholl14}, respectively. The velocity data points are taken
from \protect\cite{Nicholl15}.}
\label{fig:demo}
\end{figure*}

A comparison of Figures \ref{fig:SN-T} and \ref{fig:Te} indicates that the
temperature evolution could be slightly complicated if the optical transient
acquires a quasi-relativistic speed. In this case, the photosphere recedes
rapidly that the effective temperature could rise before the photosphere
disappears.

From Equations $\left( \ref{eq:SN-ph-lum}\right) $ and $\left( \ref%
{eq:phi_dot_photosphere}\right) $ we see that for a sub-relativistic SLSN,
at the maximum of the light curve, the SN luminosity equals the
instantaneous energy input rate \citep{Arnett79,Arnett82}, under the
assumption $x_{\mathrm{ph}}=1$ at the peak luminosity. It is clear here,
because of the photospheric recession, that this law is only a good
approximation. Assuming that the photospheric luminosity dominates over the
nebular luminosity at the peak time $t_{\mathrm{pk}}$, Equations $\left( \ref%
{eq:SN-ph-lum}\right) $ and $\left( \ref{eq:phi_dot_photosphere}\right) $
give%
\begin{equation}
\frac{dL_{\mathrm{ph}}}{dt}=\frac{E_{\mathrm{th}}\left( 0\right) }{\tau
_{0}x_{\mathrm{ph}}^{2}}\frac{R\left( t\right) }{R\left( 0\right) }\left[ 
\frac{L_{\mathrm{inp}}^{\mathrm{SN}}\left( t\right) }{E_{\mathrm{th}}\left(
0\right) }-\frac{x_{\mathrm{ph}}}{\tau _{0}}\phi -2x_{\mathrm{ph}}^{2}\frac{%
R\left( 0\right) }{R\left( t\right) }\dot{x}_{\mathrm{ph}}\phi \right] .
\end{equation}%
Because $\dot{x}_{\mathrm{ph}}<0$, at the peak time $t_{\mathrm{pk}}$, i.e. $%
dL_{\mathrm{ph}}/dt=0$, the above equation gives%
\begin{equation}
L_{\mathrm{inp}}^{\mathrm{SN}}\left( t_{\mathrm{pk}}\right) <L_{\mathrm{ph}%
}\left( t_{\mathrm{pk}}\right) .
\end{equation}%
By assumption, $L_{\mathrm{inp}}^{\mathrm{atm}}\left( t_{\mathrm{pk}}\right)
\ll L_{\mathrm{inp}}^{\mathrm{SN}}\left( t_{\mathrm{pk}}\right) $, $L_{%
\mathrm{atm}}\left( t_{\mathrm{pk}}\right) \ll L_{\mathrm{ph}}\left( t_{%
\mathrm{pk}}\right) $, in general, we have 
\begin{equation}
t_{\times }<t_{\mathrm{pk}},
\end{equation}%
where $t_{\times }$ is the time when the SN luminosity equals the
instantaneous energy input rate.

This modification to the law $t_{\times }=t_{\mathrm{pk}}$ can be understood
as follows. Although the photospheric recession reduces the emitting
surface, it increases the surface temperature as well. These two effects
result in a progressive enhancement of the photospheric luminosity, i.e. the
delay of $t_{\mathrm{pk}}$ relative to $t_{\times }$.

If the optical transient gains a relativistic speed, Doppler transformation
between the comoving frame and the observer frame may affect the above law.
In the comoving frame, by definition, one have%
\begin{equation}
L_{\mathrm{ph}}^{\prime }\left( t_{\times }^{\prime }\right) =L_{\mathrm{inp}%
}^{\prime }\left( t_{\times }^{\prime }\right) .
\end{equation}%
In the observer frame, one have%
\begin{eqnarray}
L_{\mathrm{ph}}\left( t_{\times }^{\prime }\right) &=&L_{\mathrm{ph}%
}^{\prime }\left( t_{\times }^{\prime }\right) \mathcal{D}_{\times }^{2}, \\
L_{\mathrm{ph}}\left( t_{\mathrm{pk}}^{\prime }\right) &=&L_{\mathrm{ph}%
}^{\prime }\left( t_{\mathrm{pk}}^{\prime }\right) \mathcal{D}_{\mathrm{pk}%
}^{2}.
\end{eqnarray}%
Because we always have $\mathcal{D}_{\times }>$ $\mathcal{D}_{\mathrm{pk}}$,
by demanding $t_{\times }=t_{\mathrm{pk}}$, i.e. $L_{\mathrm{ph}}\left(
t_{\times }\right) =L_{\mathrm{ph}}\left( t_{\mathrm{pk}}\right) $, the
condition%
\begin{equation}
\frac{L_{\mathrm{ph}}^{\prime }\left( t_{\times }^{\prime }\right) }{L_{%
\mathrm{ph}}^{\prime }\left( t_{\mathrm{pk}}^{\prime }\right) }=\frac{%
\mathcal{D}_{\mathrm{pk}}^{2}}{\mathcal{D}_{\times }^{2}}<1
\end{equation}%
could be fulfilled as long as the dynamic evolution of the optical transient
is appropriate. The above inference indicates that it is even possible that $%
t_{\times }>t_{\mathrm{pk}}$ under relativistic motion. Figure \ref%
{fig:luminosity} shows that the light curve of the merger-nova considered
here is in the normal case, i.e. $t_{\times }<t_{\mathrm{pk}}$, where to be
seen clearly, we depicts the horizontal axis in logarithmic scale. This is
because the Doppler factor is nearly unity during the evolution of the
merger-nova.

For completeness, we mention that nonhomologous expansion, e.g. shock
breakout, can also modify the approximate law $t_{\times }=t_{\mathrm{pk}}$.
Nonhomologous expansion adds the following effective source term %
\citep{Arnett80}%
\begin{equation}
S=-\frac{4}{3}\left( -\frac{1}{\eta }\frac{d\eta }{dx}\right) \left( \frac{%
v-x\dot{R}}{R}\right) T^{4}  \label{eq:nonhomologous-source}
\end{equation}%
to the energy conservation equation%
\begin{equation}
4T^{4}\left( \frac{\dot{T}}{T}+\frac{\dot{R}}{R}\right) =\tilde{\epsilon}+%
\frac{1}{r^{2}}\frac{\partial }{\partial r}\left( \frac{c}{3\kappa \rho }%
r^{2}\frac{\partial T^{4}}{\partial r}\right) +S.
\label{eq:SN-energy-conserv}
\end{equation}%
Here the dimensionless function $\eta \left( x\right) $\ defines the shape
of the density distribution within the SN ejecta. In Equation $\left( \ref%
{eq:SN-energy-conserv}\right) $\ the first term on the left-hand side stands
for the thermal energy, while the second term on the same side comes from
the volume expansion. The first term on the right-hand side stands for the
heating rate, the second term is the SN luminosity. We leave the discussion
about the link between the nonhomologous heating term and the SN luminosity
to the Appendix.

The term $\left( \ref{eq:nonhomologous-source}\right) $ has a cooling effect
near the ejecta surface and a heating effect in the inner region %
\citep{Arnett80}. When the photosphere recedes inwards, the cooling effect
near the surface disappears but the heating effect in the inner region
remains, resulting in an enhanced luminosity relative to the homologous
expansion. Shock breakout usually manifests itself as a luminosity spike,
while the disappearance of the cooling effect of the surface acts as a bump
in the light curve.

As an analytical model, it can be benefited from a comparison with the
numerical simulations. Figure \ref{fig:Kasen} is a comparison between this
analytical model and the one-dimensional hydrodynamical simulations of \cite%
{Kasen10}. It can be seen that the luminosity calculated by the analytical
model is higher than simulations. This is the result of the diffusion
approximation inherent in the analytical solution. As shown in this figure,
increasing the dipole magnetic field (and thence reducing the spin-down time
of the magnetar) of the magnetar results in a light curve of dimmer and
short-lived. This again reinforces our conclusion that a magnetar with short
spin-down timescale converts its rotational energy mostly into the kinetic
energy of the SN, while the one with long spin-down timescale converts its
rotational energy into the radiation of the SN. An SLSN is powered by a fast
rotating (relatively) weakly magnetized magnetar. This stimulates us to
speculate that the hypernovae, i.e. those SNe with kinetic energy $%
E_{K}>10^{52}\unit{erg}$,\ could be powered by millisecond magnetars with
much stronger dipole magnetic field.

The analytical model presented in this paper predicts a slow transition to
nebular phase and a flat velocity curve. This is in immediate accord with
the well known observational fact that SLSNe are usually slowly evolved
compared to the normal type Ic SNe \citep[e.g.][]{Nicholl15}. As a
demonstration, Figure \ref{fig:demo} shows the light curves (the
observational data are taken from \citealt{Inserra13} and \citealt{Nicholl14}%
) and velocity curves of SLSNe 2013dg and 2011ke.\footnote{%
As for 2011ke, the first luminosity data point is coincident with an
observational upper limit, as can be seen from Figure 12 in \cite{Inserra13}%
. After analyzing the data points, we think that the first data point is
quite probably an upper limit because leaving this point as an upper limit
results in a much better fit to the observational data. The data points
presented here are therefore shifted by 7.9 days, both for the luminosity
and velocity data points. As a result, of course, the fitting parameter
values are different from those given by \cite{Inserra13}.} It can be seen
that the model curves are in good agreement with the observational data. The
flatness of the velocity curves of SLSNe is the result of two factors. First
of all, SLSNe tend to be more massive than normal type Ic SNe %
\citep[e.g.][]{Nicholl15}. The magnetar, however, is another factor to make
the velocity curve flat because the magnetar keeps accelerate the ejecta.

If it were not the effect of the magnetar, the photospheric velocity would
keep declining because of the recession of the photosphere. The acceleration
of the ejecta by magnetar partially compensates the photospheric recession.
This fact may serve as a very interesting evidence in favor of the magnetar
model against other alternative models. It can be tentatively concluded that
the slow evolution to the nebular phase and the flat velocity curve of SLSNe
are the result of the large ejecta mass and the existence of magnetar
inherent to SLSNe. This gives justification for making more detailed
modelling of the magnetar-powered SLSNe.

In this model the photospheric velocity first increases rapidly and then
declines slowly. The inset of Figure \ref{fig:SN-v}, however, shows that the
photospheric velocity of SN2010gx declines more rapidly than our model. This
could be remedied by introducing an envelope as done by \cite{Inserra13}.

One issue we have deferred to discuss is the form of the energy flux from
the magnetar. In its simplest form we just assume that the rotational energy
of the magnetar is absorbed by the ejecta in the form of pure photons.
Observation of pulsar wind nebulae \citep[PWNe;][]{Gaensler06,Hester08},
however, indicates that the PWNe are usually a mixture of electron/positron
pairs and magnetic (Poynting) flux. If the ejecta are massive enough, the
rotational energy of the magnetar will eventually be completely absorbed,
regardless of the existence of electron pairs or Poynting flux. If, however,
the ejecta is less massive, e.g. $M_{\mathrm{ej}}=10^{-4}-10^{-2}M_{\odot }$%
, in the extreme case that the energy flux from the magnetar is completely
lepton dominated, the outflow ejected by the double neutron star merger
could be rapidly accelerated to relativistic speed and a strong reverse
shock develops within the ejecta \citep{wang13b,wang15}.

In most cases the PWNe behave neither in the Poynting-flux extreme, nor in
the lepton-flux extreme, but rather lie in between. We therefore expect that
the optical transients with mass as low as $M_{\mathrm{ej}%
}=10^{-4}-10^{-2}M_{\odot }$ would render their light curves as a trade-off
of that studied in this paper and that given by \cite{wang13b} and \cite%
{wang15,Wang16}, which we will study in the following papers.

\begin{acknowledgements}
We thank the referee for constructive suggestions that have allowed us to improve the
manuscript. We also thank Yun-Wei Yu for helpful discussions.
This work is supported by the National Basic Research Program (``973" Program)
 of China under Grant No. 2014CB845800 and the National Natural Science
 Foundation of China (grant Nos. U1331202, 11573014, and 11322328).
D.X. acknowledges the support of the One-Hundred-Talent Program from the
National Astronomical Observatories, Chinese Academy of Sciences.
 X.F.W. was also partially supported by 
 the Youth Innovation Promotion Association (2011231), and the Strategic
 Priority Research Program ``The Emergence of Cosmological Structure''
 (grant No. XDB09000000) of the Chinese Academy of Sciences.
\end{acknowledgements}

\appendix

\section{Basic equations and the nonhomologous source term}

\label{sec:SN-equations}

To help understand the link between Equations $\left( \ref%
{eq:nonhomologous-source}\right) $ and $\left( \ref{eq:SN-phi-dot-no-recede}%
\right) $, we briefly list the equations governing the SN evolution as
follows \citep{Arnett80,Arnett82,Arnett89}. The first law of thermodynamics
is%
\begin{equation}
\dot{E}+P\dot{V}=\epsilon -\frac{\partial L}{\partial m},
\end{equation}%
where $E$, $V$, $\epsilon $ are the thermal energy, volume, and heating rate
of unit mass, respectively. $P\left( r,t\right) $, $L\left( r,t\right) $ are
the pressure and luminosity at radius $r$, respectively. $m\left( r,t\right) 
$ is the ejecta mass within $r$. For a radiation gas, one has $E=aT^{4}V$
and $P=aT^{4}/3$. In the diffusion approximation%
\begin{equation}
\frac{L}{4\pi r^{2}}=-\frac{\lambda c}{3}\frac{\partial aT^{4}}{\partial r}
\end{equation}%
one finds that the energy conservation equation becomes%
\begin{equation}
4T^{4}\left( \frac{\dot{T}}{T}+\frac{\dot{V}}{3V}\right) =\tilde{\epsilon}+%
\frac{1}{r^{2}}\frac{\partial }{\partial r}\left( \frac{c}{3\kappa \rho }%
r^{2}\frac{\partial T^{4}}{\partial r}\right) ,
\end{equation}%
where%
\begin{equation}
\tilde{\epsilon}=\frac{\epsilon }{aV}.
\end{equation}

In the adiabatic approximation, one has $T\propto R\left( t\right) ^{-1}$.
In order to determine the temperature distribution within the ejecta, we try
to separate the dependence on space $\left( r\right) $ and time $\left(
t\right) $%
\begin{equation}
T\left( r,t\right) ^{4}=\psi \left( x\right) \phi \left( t\right) T\left(
0,0\right) ^{4}R\left( 0\right) ^{4}/R\left( t\right) ^{4},
\end{equation}%
where the dimensionless Lagrangian coordinate is defined as%
\begin{equation}
x=r/R\left( t\right) .
\end{equation}%
With such a variable separation, the energy conservation equation becomes%
\begin{equation}
\dot{\phi}+\frac{R}{R_{0}}\frac{\phi }{\tau _{0}}=\frac{b}{aT_{0}^{4}V_{0}}%
\frac{R}{R_{0}}f\left( t\right) ,
\end{equation}%
where we have isolated an (assumed) constant%
\begin{equation}
b\equiv \frac{\xi \left( x\right) \eta \left( x\right) }{\psi \left(
x\right) }
\end{equation}%
by assuming that the energy heating rate $\epsilon \left( r,t\right) $ can
be separated in space and time%
\begin{equation}
\epsilon =\xi \left( x\right) f\left( t\right)
\end{equation}%
with $\xi \left( x\right) $ a dimensionless function. With these
developments, it is easy to show that the SN luminosity is given by Equation 
$\left( \ref{eq:SN-phi-dot-no-recede}\right) $. We therefore conclude that
to include the nonhomologous expansion effect, one just needs to add $S$
into the source term $\tilde{\epsilon}$. One caveat, however, is that by
approximating $b$ a constant, we actually assume that the heating rate
distribution $\xi \left( x\right) $ is proportional to the temperature
distribution $\psi \left( x\right) $, which is usually a good approximation.
The nonhomologous source term $S$ is far from such a distribution though. At
the least, $S$ is not positive definite. Consequently the nonhomologous
effect cannot be accurately described by the simple prescription $\left( \ref%
{eq:SN-phi-dot-no-recede}\right) $.

\end{document}